
\magnification=1200
\parskip 3pt plus 1pt

\font\tit=cmr10 scaled \magstep2
\font\nom=cmr10 scaled \magstephalf
\def\res{\mathop{\rm res}}

\hbox{}
\vskip 0.0cm
\line{\hfill\hbox{LPM-95/24\ \ \ }}
\line{\hfill\hbox{RU-95-39\ \ \ \ \ \ }}
\line{\hfill\hbox{hep-th/9506136}}
\vskip 1cm
\centerline{\tit
Structure Constants and Conformal Bootstrap}
\centerline{\tit  in Liouville Field Theory}
\vskip 0.5cm
\centerline{\nom A.B.Zamolodchikov}
\vskip 0.3cm
\centerline{Department of Physics and Astronomy}
\centerline{Rutgers University}
\centerline{P.O.Box 849, Piscataway, New Jersey 08855-0849}
\centerline{and}
\centerline{L.D.Landau Institute for Theoretical Physics}
\centerline{Kosygina 2, 117334, Moscow, Russia}
\vskip 0.5cm
\centerline{\nom Al.B.Zamolodchikov}
\vskip 0.3cm
\centerline{Laboratoire de Physique Math\'ematique}
\centerline{Universit\'e Montpellier II}
\centerline{Pl.E.Bataillon, 34095 Montpellier, France}
\vskip 1cm

{\bf Abstract\hfil}

An analytic expression is proposed for the three-point function of the
exponential fields in the Liouville field theory on a sphere. In the
classical limit it coincides with what the classical Liouville theory
predicts. Using this function as the structure constant of the operator
algebra we construct the four-point function of the exponential fields and
verify numerically that it satisfies the conformal bootstrap equations, i.e.,
that the operator algebra thus defined is associative. We consider also the
Liouville reflection amplitude which follows explicitly from the structure
constants.

\vskip 2cm

{\bf 1. Introduction}

Since early 80's when the two-dimensional Liouville field theory (LFT)
was recognized [1] as the effective field theory of the 2D quantum gravity
considerable efforts has been directed at this area, especially for its
relation to the string theory [1] (see e.g. refs.[2--4]). However despite
significant progress in understanding the situation up to now the solution
to LFT is generally lacking. E.g., the structure of the operator algebra and
the correlation functions in the general case are still unknown.

Interest in LFT was renewed recently with the development of the
alternative matrix model approach to the 2D gravity [5,6]. As the result
of this new activity it was shown that LFT is able to reproduce some of
the predictions of the matrix model approach, in particular the scaling
behavior [7--9], the genus one partition functions [10] and some of the
integrated correlation functions [11--15]. We would like to emphasize
refs.[12] and [14,15] where the analytic continuation in the order of
perturbation theory has been used. This approach is conceptually close
to the way we make our guess about the structure constants.

The content below is arranged as follows. In sect.2 the LFT on a sphere is
introduced and some notations are defined. In sect.3 we propose an exact
expression for the LFT three-point function and discuss some of its properties.
The Liouville reflection amplitude is also introduced here. It should be
stressed that the arguments of this section have nothing to do with a
derivation. These are rather some motivations and we consider the expression
proposed as a guess which we try to support in the subsequent sections.
This guess appears quite natural\footnote{$^1$}{Compare (3.14) below with
the formulae for the structure constants of ``minimal CFT''[16]
and other ``rational CFT''[17] which were taken as ``architectural
prototypes'' for (3.14).} and might even be thought obvious to those
concerned with the problem. What we believe does promote it
to a step forward are the various tests performed in sects.4--7.

Sect.4 contains some calculations in the semiclassical limit. The physical
content of the Liouville reflection amplitude is discussed in sect.5 and in
sect.6 we make a check of this amplitude by means of the thermodynamic Bethe
ansatz for the high-temperature sinh-Gordon model.

The exact LFT structure constants together with an effective numerical
algorithm for the conformal block permit us to compute numerically the
four-point function of the LFT exponential fields. We perform this in sect.7
and verify that the four-point function satisfies the conformal bootstrap
equations, i.e. the necessary conditions of the associativity of the operator
algebra. Sect.8 contains some considerations about the classical Liouville
action and the related problem of accessory parameters.

\vskip 0.5cm

{\bf 2. Liouville field theory}

Local properties of the LFT are derived from the Lagrangian
density\footnote{$^2$}{It is conventional to add also another term
${Q\over 4\pi}\hat R\sqrt{\hat g}\ \phi$ to the Liouville Lagrangian
density, with ${\hat g}$ and ${\hat R}$ being arbitrary "background
metric" and associated curvature, and with the parameter $Q$ adjusted to
ensure that all physical quantities be independent of a particular
choice of this "background" (see [8,9]). However, it is always possible
to choose a specific background which is flat everywhere except for few
selected points; the above term translates then into appropriate
"boundary" terms, as in (2.7) and (2.33) below.}
$$
{\cal L}={1\over 4\pi}(\partial_a\phi)^2+\mu e^{2b\phi}
\eqno(2.1)
$$
where $b$ is the dimensionless Liouville coupling constant and the scale
parameter $\mu$ is usually called the cosmological constant. Below we often
use the complex euclidean coordinates $z=x_1+ix_2$; $\bar z=x_1-ix_2$ and
denote $\partial=\partial/\partial z$; $\bar\partial=\partial/\partial
\bar z$. The Liouville field $\phi(z,\bar z)$ is not exactly a scalar but
varies under holomorphic coordinate transformations $z\to w(z)$ as
$$
\phi(w,\bar w)=\phi(z,\bar z)-{Q\over 2}\log\left|{dw\over dz}\right|^2
\eqno(2.2)
$$
where
$$
Q=b+1/b
\eqno(2.3)
$$
The holomorphic Liouville stress tensor
$$
\eqalign{
T(z)&=-(\partial\phi)^2+Q\partial^2\phi\cr
\bar T(\bar z)&=-(\bar\partial\phi)^2+Q\bar\partial^2\phi\cr
}\eqno(2.4)
$$
ensures local conformal invariance [18] of LFT with the Liouville central
charge
$$
c_L=1+6Q^2
\eqno(2.5)
$$

To define LFT globally one has to specify boundary conditions. The LFT on a
sphere corresponds to $\phi$ defined on the whole complex plane with the
following asymptotic behavior at $|z|\to\infty$
$$
\phi(z,\bar z)=-Q\log(z\bar z)+O(1)\ \ \ \ \ {\rm at}\ \ |z|\to\infty
\eqno(2.6)
$$
For specific calculations it is sometimes convenient to set this asymptotic
behavior by considering LFT on a large disk $\Gamma$ of radius $R\to\infty$
and adding a boundary term to the Liouville action
$$
A_L={1\over 4\pi}\int_\Gamma\left[(\partial_a\phi)^2+4\pi\mu e^{2b\phi}\right]
d^2x+{Q\over\pi R}\int_{\partial\Gamma}\phi dl+2Q^2\log R
\eqno(2.7)
$$
The last constant term is introduced to make the action finite at $R\to\infty$.
This type of boundary condition is conventionally called the background charge
$-Q$ at infinity.

Exponential Liouville operators
$$
V_\alpha(x)=e^{2\alpha\phi(x)}
\eqno(2.8)
$$
are the spinless primary conformal fields of dimensions
$$
\Delta_\alpha=\alpha(Q-\alpha)
\eqno(2.9)
$$
Note that the field $V_{Q-\alpha}$ has the same dimension as (2.8). These two
fields are closely related and we shall call $V_{Q-\alpha}$ the reflection
image of $V_\alpha$ and vice versa. In particular the perturbation operator
$V_b$ in (2.1) have dimension 1 together with its reflection image $V_{1/b}$.
The case $\alpha=Q/2$ is degenerate. Here we have two primary fields
$$
V_{Q/2}(x)=e^{Q\phi(x)}
\eqno(2.10)
$$
and
$$
U_{Q/2}(x)={1\over 2}{\partial\over\partial\alpha}\left. V_\alpha(x)
\right|_{\alpha=Q/2}=\phi(x)e^{Q\phi(x)}
\eqno(2.11)
$$
of dimension $Q^2/4$. The last field (2.11) is called sometimes the puncture
operator in LFT [11].

The $n$-point function of the exponential fields on a sphere is formally
defined as a functional integral
$$
{\cal G}_{\alpha_1,\ldots,\alpha_n}(x_1,\ldots,x_n)=\int V_{\alpha_1}(x_1)
\ldots V_{\alpha_n}(x_n)e^{-A_L[\phi]}D\phi
\eqno(2.12)
$$
over the fields $\phi(x)$ with the boundary condition (2.6). Note that in the
definition (2.12) we do not divide the right hand side by the zero-point
function $Z_0$.

The scale ($\mu$) dependence of any correlation function (2.12) [7--9]
$$
{\cal G}_{\alpha_1,\ldots,\alpha_n}(x_1,\ldots,x_n)=(\pi\mu)^{(Q-\sum\alpha_i)
/b}F_{\alpha_1,\ldots,\alpha_n}(x_1,\ldots,x_n)
\eqno(2.13)
$$
(with $F_{\alpha_1,\ldots,\alpha_n}(x_1,\ldots,x_n)$ independent on $\mu$)
is easily derived from the action (2.7) and the operator product expansion
$$
\phi(z,\bar z)V_\alpha(x)=-\alpha\log|z-x|^2 V_\alpha(x)+\ldots
\eqno(2.14)
$$
Sometimes it is convenient to consider the $n$-point functions
${\cal G}^{(A)}_{\alpha_1,\ldots,\alpha_n}(x_1,\ldots,x_n)$ with fixed area
$$
A=\int e^{2b\phi}d^2x
\eqno(2.15)
$$
These observables are related to (2.12) as
$$
{\cal G}_{\alpha_1,\ldots,\alpha_n}(x_1,\ldots,x_n)=\int_0^\infty
{\cal G}_{\alpha_1,\ldots,\alpha_n}^{(A)}(x_1,\ldots,x_n)e^{-\mu A}{dA\over A}
\eqno(2.16)
$$
so that
$$
{\cal G}^{(A)}_{\alpha_1,\ldots,\alpha_n}(x_1,\ldots,x_n)=
\left({A\over\pi}\right)^{(\sum\alpha_i-Q)/b}{F_{\alpha_1,\ldots,\alpha_n}
(x_1,\ldots,x_n)\over\Gamma\left(\left(Q-\sum\alpha_i\right)/b\right)}
\eqno(2.17)
$$

Conformal invariance restricts to some extent the $x$-dependence of the
correlation functions [18]. In particular the three-point function is specified
up to an $x$-independent constant $C(\alpha_1,\alpha_2,\alpha_3)$
$$
{\cal G}_{\alpha_1,\alpha_2,\alpha_3}(x_1,x_2,x_3)=
\left|x_{12}\right|^{2\gamma_3}
\left|x_{23}\right|^{2\gamma_1}\left|x_{31}\right|^{2\gamma_2}
C(\alpha_1,\alpha_2,\alpha_3)
\eqno(2.18)
$$
where $\gamma_1=\Delta_{\alpha_1}-\Delta_{\alpha_2}-\Delta_{\alpha_3}$,
$\gamma_2=\Delta_{\alpha_2}-\Delta_{\alpha_3}-\Delta_{\alpha_1}$, $\gamma_3=
\Delta_{\alpha_3}-\Delta_{\alpha_1}-\Delta_{\alpha_2}$ and here and below we
denote $x_{ij}=x_i-x_j$. The four-point function can be reduced to a function
of only one coordinate variable, the projective invariant of the four points
$x_1,\ldots,x_4$
$$
x={x_{12}x_{34}\over x_{14}x_{32}}
\eqno(2.19)
$$
as follows
$$
{\cal G}_{\alpha_1,\ldots,\alpha_4}(x_1,\ldots,x_4)=\prod_{i<j}\left|x_{ij}
\right|^{2\gamma_{ij}}G_{\alpha_1,\alpha_2,\alpha_3,\alpha_4}(x,\bar x)
\eqno(2.20)
$$
where $\gamma_{12}=\gamma_{13}=0$, $\gamma_{14}=-2\Delta_{\alpha_1}$,
$\gamma_{24}=\Delta_{\alpha_1}+\Delta_{\alpha_3}-\Delta_{\alpha_2}-
\Delta_{\alpha_4}$, $\gamma_{34}=\Delta_{\alpha_1}+\Delta_{\alpha_2}-
\Delta_{\alpha_3}-\Delta_{\alpha_4}$ and $\gamma_{23}=\Delta_{\alpha_4}-
\Delta_{\alpha_1}-\Delta_{\alpha_2}-\Delta_{\alpha_3}$. Function
$G_{\alpha_1,\alpha_2,\alpha_3,\alpha_4}(x,\bar x)$ satisfies the following
symmetries (sometimes called the crossing symmetry relations) [18]
$$
G_{\alpha_1,\alpha_2,\alpha_3,\alpha_4}(x,\bar x)=G_{\alpha_1,\alpha_3,
\alpha_2,\alpha_4}(1-x,1-\bar x)=|x|^{-4\Delta_{\alpha_1}}
G_{\alpha_1,\alpha_4,\alpha_3,\alpha_2}(1/x,1/\bar x)
\eqno(2.21)
$$

In principle one can separate two pairs of operators in the four-point
function, say $V_{\alpha_1}(x_1)V_{\alpha_2}(x_2)$ and $V_{\alpha_3}(x_3)
V_{\alpha_4}(x_4)$ and represent ${\cal G}_{\alpha_1,\ldots,\alpha_4}(x_1,
\ldots,x_4)$ as a sum over intermediate physical states [18]. As was
established long ago [2,3] the physical LFT space of states ${\cal A}$
consists of a continuum variety of primary states corresponding to operators
$V_{\alpha}$ with
$$
\alpha={Q\over 2}+iP
\eqno(2.22)
$$
($P$ is real) and the conformal descendants of these states. The corresponding
expression for the four-point function looks as
$$
G_{\alpha_1,\alpha_2,\alpha_3,\alpha_4}(x,\bar x)=
{1\over 2}\int\limits_{-\infty}^\infty
C(\alpha_1,\alpha_2,Q/2+iP)C(\alpha_3,\alpha_4,Q/2-iP)\vert{\cal F}
(\Delta_{\alpha_i},\Delta,x)\vert^2 dP
\eqno(2.23)
$$
Here $C(\alpha_1,\alpha_2,\alpha_3)$ are the structure constants of eq.(2.18)
while ${\cal F}(\Delta_{\alpha_i},\Delta,x)$ is the so-called conformal block
[18] which sums up all the intermediate descendant states of a given
primary one. The conformal block is determined completely by the conformal
symmetry of the theory. It depends on the corresponding central charge (2.5),
on the dimension
$$
\Delta={Q^2\over 4}+P^2
\eqno(2.24)
$$
of the intermediate primary state and also on the ``external'' dimensions
$\Delta_{\alpha_i}$, $i=1,\ldots,4$. Unfortunately up to now this function
is not known in a closed form. However it is straightforward to evaluate it
as a power series in $x$ [18]. Also there are very efficient algorithms for
its numerical computation [19].

One important remark about Eq.(2.23) is in order. As we will see
(sect.3) the structure constant $C(\alpha_1, \alpha_2, \alpha)$ exhibits
infinitely many poles in the variable $\alpha$, their positions being
dependent on $\alpha_1, \alpha_2$. As one changes $\alpha_1, \alpha_2$
and $\alpha_3, \alpha_4$ in (2.23) the associated poles of the integrand
move around and some of them can cross the integration contour ${\Im m}
P = 0$. Therefore the decomposition (2.23) can be taken literally only if
$\alpha_1, \alpha_2$ and $\alpha_3, \alpha_4$ take their values within
certain domains, namely
$$
\eqalign{
&|\alpha_1 - \alpha_2| < Q/2;\qquad |Q - \alpha_1 - \alpha_2| < Q/2;\cr
&|\alpha_3 - \alpha_4| < Q/2;\qquad |Q - \alpha_3 - \alpha_4| < Q/2.
}\eqno(2.25)
$$
(here we consider the case of real $\alpha's$). Otherwise the four-point
function (2.23) has to be understood as analytic continuation away from
the domain (2.25). In the course of this continuation some
poles of $C(\alpha_1, \alpha_2, Q/2+iP)$ and/or of $C(\alpha_3, \alpha_4,
Q/2+iP)$ cross the real $P$ axis and the r.h.s. of (2.23) acquires
additional ``discrete'' terms associated with the residues of the
integrand at these poles (this seems to agree with qualitative
analysis in [11]). We will discuss this phenomenon in greater
details elsewhere.

Of course one can pick up another partition of the four operators into pairs
and obtain an expression similar to (2.23) summing up over the intermediate
states in the corresponding channel. The resulting four-point function must be
the same. In other words the four-point function defined as in eq.(2.23) has
to satisfy the crossing symmetry relations (2.21). From this point of view
these relations appear as a set of non-trivial restrictions for the structure
constants $C(\alpha_1,\alpha_2,\alpha_3)$ which express the associativity of
the operator algebra and are known as the conformal bootstrap equations [18].
We discuss more about this point in sect.7. Representations like (2.23) can
be written down for higher multipoint functions as well. However they involve
the multipoint conformal blocks which are much more complicated objects then
the four-point one. Fortunately, the corresponding crossing symmetry
relations are not expected to bring up any new restrictions on the structure
of the operator algebra [18].

Let us finish this section with few words about the classical limit $b\to 0$
of LFT. Here it is more convenient to use the field
$$
\varphi=2b\phi
\eqno(2.26)
$$
which becomes a classical Liouville field in this limit. Its dynamics is
governed by the classical action
$$
S_{\rm Liouv}[\varphi]=b^2 A_L[\phi]\ \ \ \ \ {\rm at}\ \ b\to 0
\eqno(2.27)
$$
Explicitly
$$
S_{\rm Liouv}[\varphi]={1\over 8\pi}\int_\Gamma\left[{1\over 2}(\partial_a
\varphi)^2+8\pi\mu b^2e^\varphi\right]d^2x+\varphi_\infty+2\log R
\eqno(2.28)
$$
where
$$
\varphi_\infty={1\over 2\pi R}\int_{\partial\Gamma}\varphi dl
\eqno(2.29)
$$
The field $\varphi(x)$ satisfies the classical Liouville equation
$$
\partial\bar\partial\varphi=2\pi\mu b^2e^\varphi
\eqno(2.30)
$$
and locally describes a surface of constant negative curvature $-8\pi\mu b^2$.

The leading (exponential) asymptotic of the $n$-point function (2.12) in the
classical limit
$$
{\cal G}_{\alpha_1,\ldots,\alpha_n}(x_1,\ldots,x_n)\sim\exp\left(
-{1\over b^2}S^{({\rm cl})}\right)
\eqno(2.31)
$$
is governed by the classical Liouville action on an appropriate solution to
the Liouville equation. Note that the insertion of any $V_{\alpha_i}(x_i)$
in (2.12) affects the classical field dynamics only if the corresponding
$\alpha_i$ is ``heavy'', i.e. if $\alpha_i=\eta_i/b$ is of order $O(b^{-1})$.
Technically one has to distinguish two possibilities. If $\sum_i\eta_i>1$ a
classical solution with negative curvature exists and in (2.30)
$$
S^{({\rm cl})}_{\eta_1,\ldots,\eta_n}(x_1,\ldots,x_n)=
S_{\rm Liouv}\left[\varphi_{\eta_1,\ldots,\eta_n}(x|x_1,\ldots,x_n)\right]
\eqno(2.32)
$$
where $\varphi_{\eta_1,\ldots,\eta_n}(x|x_1,\ldots,x_n)$ is a solution to
(2.30) with the following boundary conditions
$$
\eqalign{
\varphi(z,\bar z)&=-2\log|z|^2+O(1)\ \ \ \ \ \ \
\ \ \ \ \ \ \ \ {\rm at}\ \ |z|\to\infty\cr
\varphi(z,\bar z)&=-2\eta_i\log|z-x_i|^2+O(1)\ \ \ \ \ \
{\rm at}\ \ z\to x_i
}\eqno(2.33)
$$
This field configuration is singular at $z\to x_i$. Therefore it is better
to cut out small disks of radius $\epsilon_i$ around each point $x_i$ and
define the regularized Liouville action on the remaining part $\Gamma$ of the
complex plane
$$
S_{\rm Liouv}[\varphi]={1\over 8\pi}\int_\Gamma\left[{1\over 2}(\partial_a
\varphi)^2+8\pi\mu b^2e^\varphi\right]d^2x+\varphi_\infty+2\log R-
\sum_i\left(\eta_i\varphi_i+2\eta_i^2\log\epsilon_i\right)
\eqno(2.34)
$$
Here $\varphi_\infty$ is as in eq.(2.29) while the boundary terms with
$$
\varphi_i={1\over 2\pi\epsilon_i}\int_{\partial\Gamma_i}\varphi dl
\eqno(2.35)
$$
are added for each small circle to ensure the behavior (2.33) near $x_i$.
Also we include some field independent terms such that $S_{\rm Liouv}$ is
finite and independent on $\epsilon_i$ at $\epsilon_i\to 0$.

At $\sum\eta_i<1$ there is no solution to (2.30) and (2.33) with negative
curvature. In this case it is relevant to consider the $n$-point function
(2.17) of fixed area. The leading classical behavior is again governed by
the classical action
$$
{\cal G}^{(A)}_{\alpha_1,\ldots,\alpha_n}(x_1,\ldots,x_n)\sim
\exp\left(-{1\over b^2}S_A^{({\rm cl})}\right)
\eqno(2.36)
$$
but now one has to impose the constant area constraint
$$
A=\int_\Gamma e^\varphi d^2x
\eqno(2.37)
$$
and solve the positive curvature Liouville equation
$$
\partial\bar\partial\varphi={2\pi\over A}\left(\sum\eta_i-1\right)e^\varphi
\eqno(2.38)
$$
with the same boundary conditions (2.33). The corresponding constant area
Liouville action is again (2.34) but without the cosmological term
$\mu b^2 A$ so that
$$
S_{\rm Liouv}[\varphi]=S^{({\rm cl})}_A[\varphi]+\mu b^2\int_\Gamma
e^\varphi d^2x
\eqno(2.39)
$$
Note that the same fixed area calculation can be performed in the first case
$\sum\eta_i>1$ too. The integral over $A$ in eq.(2.16) is now dominated by
the stationary point $A=(\sum\eta_i-1)/(\mu b^2)$ which corresponds to
the classical solution of eq.(2.30).

Suppose now that the multipoint function contains several ``heavy'' operators
with $\alpha_i=\eta_i/b$ and also a number of ``light'' ones, i.e. the
Liouville exponentials with $\alpha_j=\sigma_j b$ of the order $O(b)$. In the
classical limit these ``light'' fields influence neither the classical
solution nor the one-loop correction. Therefore one expects that as $b\to 0$
$$
{{\cal G}_{\sigma_1b,\ldots,\sigma_lb,\eta_1/b,\ldots,\eta_n/b}(y_1,\ldots,y_l,
x_1,\ldots,x_n)\over{\cal G}_{\eta_1/b,\ldots,\eta_n/b}(x_1,\ldots,x_n)}=
\prod_{j=1}^l e^{\sigma_j\varphi_{\eta_1,\ldots,\eta_l}(y_j|x_1,\ldots,x_n)}
\eqno(2.40)
$$
where $\varphi_{\eta_1,\ldots,\eta_l}(x|x_1,\ldots,x_n)$ is the classical
solution for the ``heavy'' operator configuration. Special care is required
if the number of the ``heavy'' operators is less then 3. In this case the
functional integral (2.12) has zero modes which have to be integrated out
explicitly even in the classical limit. In particular, if there are no
``heavy'' operators at all the relevant fixed area classical solution is the
2D sphere metric of area $A$
$$
\varphi_0(z,\bar z)=\log{A\over\pi(1+z\bar z)^2}
\eqno(2.41)
$$
This solution has to be integrated over its $SL(2,C)$ orbit parameterized by
four complex numbers $a$, $b$, $c$, $d$ with the constraint $ad-bc=1$
$$
\varphi_0(z,\bar z|a,b,c,d)=\log{A\over\pi(|az+b|^2+|cz+d|^2)^2}
\eqno(2.42)
$$
This leads to the following expression for the $n$-point function of ``light''
operators in the classical limit [20]
$$
{{\cal G}^{(A)}_{\sigma_1b,\ldots,\sigma_nb}(x_1,\ldots,x_n)\over
Z_0^{(A)}}=\int\prod_{i=1}^n e^{\sigma_i\varphi_0(x_i,\bar x_i|a,b,c,d)}
d\mu (a,b,c,d)
\eqno(2.43)
$$
where $Z_0^{(A)}$ is the fixed area zero-point function (the partition
function of the sphere) and $d\mu(a,b,c,d)$ stands for the invariant measure
on $SL(2,C)$,
$$
d\mu(a,b,c,d)=4d^2a\ d^2b\ d^2c\ d^2d\ \delta^{(2)}(ad-bc-1).
\eqno(2.44)
$$

\vskip 0.5cm

{\bf 3. Three-point function}

Naively one can try to expand the $N$-point function (2.12) into a
perturbative series in the cosmological constant $\mu$
$$
{\cal G}_{\alpha_1,\ldots,\alpha_N}(x_1,\ldots,x_N)=
\sum_{n=0}^\infty {\cal G}_{\alpha_1,\ldots,\alpha_N}^{(n)}(x_1,\ldots,x_N)
\eqno(3.1)
$$
where
$$
{\cal G}_{\alpha_1,\ldots,\alpha_N}^{(n)}={(-\mu)^n\over n!}
\int\left<V_{\alpha_1}(x_1)\ldots V_{\alpha_N}(x_N)V_b(u_1)\ldots V_b(u_n)
\right>d^2u_1\ldots d^2u_n
\eqno(3.2)
$$
and $\left<\ldots\right>$ denotes the functional integral over a free field
(i.e., (2.12) at $\mu=0$)
$$
\eqalign{
\left<V_{\alpha_1}(x_1)\ldots V_{\alpha_N}(x_N)\right>&=
\int\prod_{i=1}^N e^{2\alpha_i\phi(x_i)}
\exp\left(-{1\over 4\pi}\int(\partial_a\phi)^2
d^2x\right)D\phi\cr
&=\prod_{i>j}\left|x_i-x_j\right|^{-4\alpha_i\alpha_j}\cr
}\eqno(3.3)
$$
However it is well known that in LFT expression (3.1) in general does not
work. The problem is that the free field functional integral (3.3) for the
$n$-th term (3.2) matches the spherical boundary condition (2.6) only if
$$
\sum_{i=1}^N\alpha_i=Q-nb
\eqno(3.4)
$$
Following refs.[12--14] we shell interpret eq.(3.4) as a kind of
``on-mass-shell'' condition. Namely, ${\cal G}_{\alpha_1,\ldots,
\alpha_N}(x_1,\ldots,x_N)$ exhibit a pole in the variable $\alpha=\sum\alpha_i$
every time eq.(3.4) is satisfied for $n=0,1,2,\ldots$, the residue being
specified by the corresponding perturbative integral
$$
\res_{\sum\alpha_i=Q-nb}{\cal G}_{\alpha_1,\ldots,\alpha_N}(x_1,\ldots,x_N)
=\left.{\cal G}^{(n)}_{\alpha_1,\ldots,\alpha_N}(x_1,\ldots,x_N)
\right|_{\sum\alpha_i=Q-nb}
\eqno(3.5)
$$

It seems unlikely that the on-mass-shell condition (3.5) alone is enough
to determine the $N$-point function. The analysis is hampered by the quite
complicated and in general unknown analytic structure of the multipoint
perturbative integrals (3.2). The situation is simplified in the three-point
case $N=3$. The $x$-dependence of the on-mass-shell integrals turns out to
be the same as that of the conformal three-point function (2.18)
$$
\left.{\cal G}^{(n)}_{\alpha_1,\alpha_2,\alpha_3}(x_1,x_2,x_3)
\right|_{\sum\alpha_i=
Q-nb}=\left|x_{12}\right|^{2\gamma_3}\left|x_{23}\right|^{2\gamma_1}
\left|x_{31}\right|^{2\gamma_2}I_n(\alpha_1,\alpha_2,\alpha_3)
\eqno(3.6)
$$
with the same $\gamma_1$, $\gamma_2$ and $\gamma_3$ as in eq.(2.18), while
$I_n(\alpha_1,\alpha_2,\alpha_3)$ has been carried out explicitly in
ref.[21]
$$
I_n(\alpha_1,\alpha_2,\alpha_3)=\left({-\pi\mu\over\gamma(-b^2)}\right)^n
{\prod_{j=1}^n\gamma(-jb^2)\over\prod_{k=0}^{n-1}\left[\gamma(2\alpha_1b+kb^2)
\gamma(2\alpha_2b+kb^2)\gamma(2\alpha_3b+kb^2)\right]}
\eqno(3.7)
$$
Here and below the standard notation
$$
\gamma(x)=\Gamma(x)/\Gamma(1-x)
\eqno(3.8)
$$
is used. The on-mass-shell condition now reads
$$
\res_{\sum\alpha_i=Q-nb}C(\alpha_1,\alpha_2,\alpha_3)=I_n(\alpha_1,\alpha_2,
\alpha_3)
\eqno(3.9)
$$

At this step we need to introduce some special function
$\Upsilon(b,x)$. Below we
consider $b$ as a parameter and suppress it in the notation. In the strip
$0<{\rm Re}x<Q$ function $\Upsilon(x)$ is defined by the integral
representation
$$
\log\Upsilon(x)=
\int_0^\infty{dt\over t}\left[\left({Q\over 2}-x\right)^2e^{-t}
-{\sinh^2\left({Q\over 2}-x\right){t\over 2}\over\displaystyle\sinh{bt\over 2}
\sinh{t\over 2b}}\right]
\eqno(3.10)
$$
{}From the definition it is clear that
$$
\eqalign{
\Upsilon(x)&=\Upsilon(Q-x)\cr
\Upsilon(Q/2)&=1\cr
}\eqno(3.11)
$$
and that $\Upsilon(x)$ is ``self-dual'', i.e. remains unchanged if
$b\to 1/b$. Also the following functional relations are derived
$$
\eqalignno{
\Upsilon(x+b)&=\gamma(bx)b^{1-2bx}\Upsilon(x)&(3.12a)\cr
\Upsilon(x+1/b)&=\gamma(x/b)b^{2x/b-1}\Upsilon(x)&(3.12b)\cr
}
$$
Using (3.10) and (3.12) it is easy to verify that $\Upsilon(x)$ is an entire
function of $x$ with zeroes located at $x=-m/b-nb$ and $x=(m+1)/b+(n+1)b$,
where $m$ and $n$ run over all non-negative integers. Below we also use the
notation
$$
\Upsilon_0=\left.{d\Upsilon(x)\over dx}\right|_{x=0}
\eqno(3.13)
$$

With the relations (3.12) it is straightforward to verify that
$$
\eqalign{
C(\alpha_1,\alpha_2,\alpha_3)&=\left[\pi\mu\gamma(b^2)b^{2-2b^2}
\right]^{(Q-\sum\alpha_i)/b}\times\cr
&\phantom{Xx}
{\Upsilon_0\Upsilon(2\alpha_1)\Upsilon(2\alpha_2)\Upsilon(2\alpha_3)\over
\Upsilon(\alpha_1+\alpha_2+\alpha_3-Q)
\Upsilon(\alpha_1+\alpha_2-\alpha_3)\Upsilon(\alpha_2+\alpha_3-\alpha_1)
\Upsilon(\alpha_3+\alpha_1-\alpha_2)}
}\eqno(3.14)
$$
satisfies the on-mass-shell condition (3.9). We propose this expression as the
exact three-point function in LFT.

As a function of $\alpha=\sum_{i=1}^3\alpha_i$ expression (3.14) has more
poles then predicted by eq.(3.9). They appear at $\alpha=Q-m/b-nb$ and
at $\alpha=2Q+m/b+nb$ for any pair of non-negative integers $m$ and $n$. It
seems suggestive to note that the corresponding residues are related to
the more general multiple integrals, also evaluated explicitly in ref.[21].
Namely
$$
\eqalign{
&\res_{\sum\alpha_i=Q-m/b-nb}{\cal
G}_{\alpha_1,\alpha_2,\alpha_3}(x_1,x_2,x_3)=
\cr
&{(-\tilde\mu)^m(-\mu)^n\over m!n!}\int\left<\prod_{i=1}^3 V_{\alpha_i}(x_i)
\prod_{j=1}^m V_{1/b}(u_j)\prod_{j=1}^n V_b(v_j)\right>d^2u_1\ldots d^2u_m
d^2v_1\ldots d^2v_n\cr
}\eqno(3.15)
$$
where the ``dual'' cosmological constant $\tilde\mu$ is related to $\mu$ as
$$
\pi\tilde\mu\gamma(1/b^2)=\left(\pi\mu\gamma(b^2)\right)^{1/b^2}
\eqno(3.16)
$$
The whole expression (3.14) is self-dual in the sense that it is invariant
under the substitution $b\to 1/b$, $\mu\to\tilde\mu$.

The three-point function (3.14) reveals some interesting features. In
particular the reflection $\alpha\to Q-\alpha$ of each of the three operators
introduces the Liouville reflection amplitude $S(P)$
$$
C(Q-\alpha_1,\alpha_2,\alpha_3)=C(\alpha_1,\alpha_2,\alpha_3)S(i\alpha_1-iQ/2)
\eqno(3.17)
$$
which reads explicitly
$$
S(P)=-\left(\pi\mu\gamma(b^2)\right)^{-2iP/b}
{\Gamma(1+2iP/b)\Gamma(1+2iPb)\over\Gamma(1-2iP/b)\Gamma(1-2iPb)}
\eqno(3.18)
$$
This function is discussed more in sect.5. We shall also use the notation
$$
G(\alpha)=S(-i\alpha+iQ/2)={\displaystyle\left(\pi\mu\gamma(b^2)\right)^{(Q-
2\alpha)/b}\over b^2}{\gamma(2\alpha b-b^2)\over\displaystyle\gamma\left(
2-2\alpha/b+1/b^2\right)}
\eqno(3.19)
$$
associating $G(\alpha)$ with the two-point function of operators $V_\alpha(x_1)
V_\alpha(x_2)$ in LFT.

There are several different classical limits of the three-point function (3.14)
dependent on how the parameters $\alpha_i$ behave as $b\to 0$. We consider
here two cases. First, let all the three operators be ``heavy'', i.e.
$\alpha_i=\eta_i/b$ at $b\to 0$ with $\eta_i$ fixed. The leading asymptotic
is
$$
C(\alpha_1,\alpha_2,\alpha_3)\sim\exp\left({1\over b^2}S^{({\rm cl})}
(\eta_1,\eta_2,\eta_3)\right)
\eqno(3.20)
$$
where
$$
\eqalign{
S^{({\rm cl})}(\eta_1,\eta_2,\eta_3)&=(\sum\eta_i-1)\log(\pi\mu b^2)+
F(\eta_1+\eta_2+\eta_3-1)+F(\eta_1+\eta_2-\eta_3)+\cr
&\phantom{Xx}F(\eta_2+\eta_3-\eta_1)+F(\eta_3+\eta_1-\eta_2)-
F(0)-F(2\eta_1)-F(2\eta_2)-F(2\eta_3)\cr
}\eqno(3.21)
$$
and we have denoted
$$
F(\eta)=\int_{1/2}^\eta\log\gamma(x)dx
\eqno(3.22)
$$
Note that $S^{({\rm cl})}(\eta_1,\eta_2,\eta_3)$ vanishes if $\sum_{i=1}^3
\eta_i=1$.

In the opposite case of three ``light'' exponentials with $\alpha_i=b\sigma_i$
of the order $O(b)$ it is more relevant to consider the fixed area three-point
function (2.17). For this we find at $b\to 0$
$$
\eqalign{
C^{(A)}(b\sigma_1,b\sigma_2,b\sigma_3)&=\left({A\over\pi}\right)^{\sum\sigma_i
-1-1/b^2}{e^{1/b^2}e^{-2C}\over\sqrt{2\pi}b^4}\times\cr
&{\Gamma(\sigma_1+\sigma_2+\sigma_3-1)\Gamma(\sigma_1+\sigma_2-\sigma_3)
\Gamma(\sigma_2+\sigma_3-\sigma_1)\Gamma(\sigma_3+\sigma_1-\sigma_2)\over
\Gamma(2\sigma_1)\Gamma(2\sigma_2)\Gamma(2\sigma_3)}\cr
}\eqno(3.23)
$$
where $C$ is the Euler's constant.

We present also the classical limit of the fixed area two-point function
$G^{(A)}(\alpha)$, related to (3.19) as in eq.(2.16). In the ``heavy'' limit
$\alpha=\eta/b$, $b\to 0$ the leading exponential asymptotic
$$
G^{(A)}(\eta/b)\sim\exp\left(-{1\over b^2}S^{({\rm cl})}(\eta)\right)
\eqno(3.24)
$$
is governed by the classical action
$$
S^{({\rm cl})}(\eta)=(1-2\eta)\left(\log{A\over\pi}+\log(1-2\eta)-1\right)
\eqno(3.25)
$$
The fixed area two-point function of two ``light'' operators with $\alpha=
b\sigma$ reads at $b\to 0$
$$
G^{(A)}(b\sigma)=\left({A\over\pi}\right)^{2\sigma-1-1/b^2}{e^{1/b^2}e^{-2C}
\over\sqrt{2\pi}b^3(2\sigma-1)}
\eqno(3.26)
$$

\vskip 0.5cm

{\bf 4. Classical limit}

The asymptotic (3.20), (3.21) has to be compared with the classical
action (2.34) evaluated at the solution $\varphi_{\eta_1,\eta_2,\eta_3}
(z|x_1,x_2,x_3)$ of the classical Liouville equation (2.30) with three
singular points $x_1$, $x_2$ and $x_3$. The boundary conditions (2.33)
at these points are automatically imposed by the boundary terms in (2.34).
In the three-point case the solution $\varphi_{\eta_1,\eta_2,\eta_3}
(z|x_1,x_2,x_3)$ can be found explicitly in terms of hypergeometric functions
$$
\varphi_{\eta_1,\eta_2,\eta_3}(z|x_1,x_2,x_3)=-2\log\left[a_1\psi_1(z)
\psi_1(\bar z)+a_2\psi_2(z)\psi_2(\bar z)\right]
\eqno(4.1)
$$
Here
$$
\eqalign{
\psi_1(z)&=(z-x_1)^{\eta_1}(z-x_2)^{1-\eta_1-\eta_3}(z-x_3)^{\eta_3}
\hbox{}_2F_1(\eta_1+\eta_3-\eta_2,\eta_1+\eta_2+\eta_3-1,2\eta_1,x)\cr
\psi_2(z)&=(z-x_1)^{1-\eta_1}(z-x_2)^{\eta_1+\eta_3-1}(z-x_3)^{1-\eta_3}
\times\cr
&\phantom{XXXXXXXXXXXXXXX}
\hbox{}_2F_1(1+\eta_2-\eta_1-\eta_3,2-\eta_1-\eta_2-\eta_3,2-2\eta_1,x)\cr
}\eqno(4.2)
$$
and we have denoted
$$
x={(z-x_1)x_{32}\over(z-x_2)x_{31}}
\eqno(4.3)
$$
while the constants $a_1$ and $a_2$ read explicitly
$$
\eqalign{
a_1^2&={\pi\mu b^2\over\left|x_{13}\right|^{4\eta_3+4\eta_1-2}\left|x_{12}
\right|^{2-4\eta_2}\left|x_{23}\right|^{2-4\eta_1}}\times\cr
&\phantom{XXXXXXXXX}
{\gamma(\eta_1+\eta_2+\eta_3-1)\gamma(\eta_1+\eta_3-\eta_2)\gamma(\eta_1+\eta_2
-\eta_3)\over\gamma^2(2\eta_1)\gamma(\eta_2+\eta_3-\eta_1)}\cr
}\eqno(4.4)
$$
and
$$
a_2=-{\pi\mu b^2\over\left|x_{13}\right|^2(1-2\eta_1)^2a_1}
\eqno(4.5)
$$

Now we have to evaluate the integral (2.34). Calculation of the Liouville
action can be simplified by the following trick. Since $\varphi_{\eta_1,\ldots,
\eta_n}(x|x_1,\ldots,x_n)$ is an extremum of (2.34) we have
$$
{\partial\over\partial\eta_i}S_{\rm Liouv}\left[\varphi_{\eta_1,\ldots,\eta_n}
(x|x_1,\ldots,x_n)\right]=-\varphi_i-4\eta_i\log\epsilon_i
\eqno(4.6)
$$
where $\varphi_i$ is defined in eq.(2.35). Near the singular point $x_i$
the solution behaves as
$$
\varphi_{\eta_1,\ldots,\eta_n}(z|x_1,\ldots,x_n)=-2\eta_i\log|z-x_i|^2+X_i+
O\left(|z-x_i|^{2-4\eta_i}\right)
\eqno(4.7)
$$
(we suppose here that all $\eta_i<1/2$). Therefore in the limit $\epsilon_i\to
0$
$$
{\partial\over\partial\eta_i}S_{\rm Liouv}=-X_i
\eqno(4.8)
$$
This equation implies that the form
$$
dS_{\rm Liouv}=-\sum_{i=1}^n X_id\eta_i
\eqno(4.9)
$$
can be integrated, defining $S_{\rm Liouv}$ up to a constant independent on
$\eta_i$. To fix this integration ambiguity we note that the Liouville action
(2.34) can be explicitly evaluated if $\sum_{i=1}^n\eta_i=1$
$$
\left.S_{\rm
Liouv}\right|_{\sum\eta_i=1}=\sum_{i<j}2\eta_i\eta_j\log|x_i-x_j|^2
\eqno(4.10)
$$

In the three-point case the coefficients $X_i$ are easily derived from the
explicit solution (4.1--5). E.g.
$$
\eqalign{
X_1&=-(1-2\eta_1)\log\left|{x_{12}x_{13}\over x_{23}}\right|^2-
\log\pi\mu b^2-\cr
&\phantom{XXXXXXXXX}
\log{\gamma(\eta_1+\eta_2+\eta_3-1)\gamma(\eta_1+\eta_2-\eta_3)\gamma(\eta_1+
\eta_3-\eta_2)\over\gamma^2(2\eta_1)\gamma(\eta_2+\eta_3-\eta_1)}\cr
}\eqno(4.11)
$$
while $X_2$ and $X_3$ are obtained from (4.11) by obvious permutations of
$\eta_i$ and $x_i$. Integrating the form (4.9) we find
$$
\eqalign{
S_{\rm Liouv}&=(\sum\eta_1-1)\log\pi\mu b^2+
(\delta_1+\delta_2-\delta_3)\log|x_{12}|^2+
(\delta_2+\delta_3-\delta_1)\log|x_{23}|^2+\cr
&\phantom{Xx}
(\delta_3+\delta_1-\delta_2)\log|x_{13}|^2+F(\eta_1+\eta_2+\eta_3-1)+
F(\eta_1+\eta_2-\eta_3)+\cr
&\phantom{Xx}F(\eta_2+\eta_3-\eta_1)+
F(\eta_3+\eta_1-\eta_2)-F(2\eta_1)-F(2\eta_2)-F(2\eta_3)-F_0(x_i)\cr
}\eqno(4.12)
$$
where $\delta_i=\eta_i(1-\eta_i)$, function $F(\eta)$ is defined by eq.(3.22)
and $F_0(x_i)$ is the integration constant independent on $\eta_i$. Comparing
with (4.10) we find $F_0(x_i)=F(0)$ in complete agreement with the asymptotic
(3.21) of the proposed exact three-point function.

Similar and even more simple calculations with the positive curvature
Liouville equation (2.37) support the classical limit (3.25) of the
two-point function.

One can also check that in the case of three ``light'' exponentials
asymptotic form (3.23) of the proposed three-point function is in
agreement with explicit semiclassical calculation through the equation
(2.43). In the case $n=3$ the $x$-dependence of the integral in (2.43)
can be easily isolated with the help of its $SL(2,C)$ transformation
properties, so that this integral takes the form
$$
|x_{12}|^{2\nu_3}|x_{23}|^{2\nu_1}|x_{31}|^{2\nu_2}\
I(\sigma_1,\sigma_2,\sigma_3),
\eqno(4.13)
$$
where $\nu_1 = \sigma_1 -\sigma_2 -\sigma_3$, $\nu_2 = \sigma_2 -\sigma_3
-\sigma_1$, $\nu_3 = \sigma_3 -\sigma_1 -\sigma_2$, and
$$
\eqalign{
&I(\sigma_1, \sigma_2, \sigma_3) =
\left({A\over\pi}\right)^{\sigma_1 + \sigma_2 + \sigma_3}\times
{\tilde I}(\sigma_1, \sigma_2, \sigma_3); \cr
&{\tilde I}(\sigma_1, \sigma_2, \sigma_3)=
\int (|b|^2 +
|d|^2)^{-2\sigma_1} (|a+b|^2 + |c+d|^2)^{-2\sigma_2} (|a|^2 +
|b|^2)^{-2\sigma_3} d\mu (a,b,c,d).
}
\eqno(4.14)
$$
It is this factor $I(\sigma_1, \sigma_2, \sigma_3)$ which, being
multiplied by the partition function $Z_{0}^{(A)}$ (see (2.43)),
is to agree with the asymptotic formula (3.23). To evaluate the integral
${\tilde I}$ in (4.14) it is convenient to use complex
coordinates $\xi_1, \xi_2, \xi_3$ on the group manifold of $SL(2,C)$
related to $a, b, c, d$ as
$$
\xi_1 = {b\over d};\quad \xi_2 = {{a+b}\over{c+d}}; \quad \xi_3 =
{a\over c} .
\eqno(4.15)
$$
In this coordinates the invariant measure on $SL(2,C)$ takes the well known
form
$$
d\mu (a,b,c,d) = {{d^2\xi_1 d^2\xi_2
d^2\xi_3}\over {|(\xi_1 - \xi_2)(\xi_2 - \xi_3)(\xi_3 - \xi_1)|^2}},
\eqno(4.16)
$$
and the integral in (4.14) simplifies as
$$
\eqalign{
\tilde I(\sigma_1, \sigma_2,& \sigma_3)\ =
\int d^2 \xi_1 d^2 \xi_2 d^2 \xi_3\cr
&|\xi_{12}|^{-2-2\nu_3}|\xi_{23}|^{-2-2\nu_1}|\xi_{31}|^{-2-2\nu_2}
(1+|\xi_1|^2)^{-2\sigma_1}(1+|\xi_2|^2)^{-2\sigma_2}
(1+|\xi_3|^2)^{-2\sigma_3}.
}
\eqno(4.17)
$$
Now, it is straightforward to verify that this integral is invariant
under the $SU(2)$ subgroup of $SL(2,C)$. Namely, the form of this
integral does not change if one substitutes
$$
\xi_i \to {{a\xi_i + b}\over{-{\bar b}\xi_i + {\bar a}}}, \qquad |a|^2 +
|b|^2 =1.
\eqno(4.18)
$$
Using this symmetry one can set, say, $\xi_3 = \infty$ and write the
integral (4.17) as
$$
\tilde I(\sigma_1, \sigma_2, \sigma_3)= \pi \int d^2\xi_1 d^2\xi_2 |\xi_1 -
\xi_2|^{-2-2\nu_3}(1+|\xi_1|^2)^{-2\sigma_1}(1+|\xi_2|^2)^{-2\sigma_2}.
\eqno(4.19)
$$
This integral can be evaluated explicitly with the result
$$
\tilde I(\sigma_1, \sigma_2, \sigma_3) = \pi^3 {{\Gamma(\sigma_1 + \sigma_2 +
\sigma_3 -1)\Gamma(\sigma_1 + \sigma_2 - \sigma_3)\Gamma(\sigma_2 +
\sigma_3 - \sigma_1)\Gamma(\sigma_3 + \sigma_1 - \sigma_2)}\over
{\Gamma(2\sigma_1)\Gamma(2\sigma_2)\Gamma(2\sigma_3)}},
\eqno(4.20)
$$
in agreement with (3.23).

\vskip 0.5cm

{\bf 5. Reflection amplitude}

Consider LFT on an infinite flat cylinder of circumference $2\pi$ with the
cartesian coordinates $x_1$, $x_2$ (as before $z=x_1+ix_2$, $\bar z=x_1-
ix_2$) and let us interpret the coordinate $x_2$ along the cylinder as
the (imaginary) time while $x_1\sim x_1+2\pi$ be the space coordinate. The
holomorphic Liouville stress tensor (2.4) allows one to construct two copies
(right and left) of Virasoro algebra $Vir$ and $\bar{Vir}$ with the
central charge (2.5)
$$
\eqalign{
\left[L_m,L_n\right]=(m-n)L_{m+n}+{c_L\over 12}(m^3-m)\delta_{m+n}\cr
\left[\bar L_m,\bar L_n\right]=(m-n)\bar L_{m+n}+{c_L\over 12}(m^3-m)
\delta_{m+n}\cr
}\eqno(5.1)
$$
where the operators $L_n$ and $\bar L_n$ appear in the expansion of the
stress tensor
$$
\eqalign{
T(z)&=-{c_L\over 24}-\sum_{n=-\infty}^\infty L_n e^{inz}\cr
\bar T(\bar z)&
=-{c_L\over 24}-\sum_{n=-\infty}^\infty \bar L_n e^{-in\bar z}\cr
}\eqno(5.2)
$$
and act in the space of states ${\cal A}$ of LFT on the constant time circle
$x_2={\rm const}$. In particular the Hamiltonian
$$
H=-{c_L\over 12}+L_0+\bar L_0
\eqno(5.3)
$$
generates translations along the time $x_2$.

The space of states ${\cal A}$ is classified in the highest weight
representations of $Vir\otimes\bar{Vir}$
$$
{\cal A}=\oplus_P{\cal A}_P
\eqno(5.4)
$$
Each conformal class ${\cal A}_P$ contains a primary state $v_P$ which
satisfies
$$
\eqalign{
L_n v_P&=\bar L_n v_P=0\ \ \ \ \ \ \ \ \ \ \ \ \ \ \ \ \ \ \ \
{\rm at}\ \ \ n>0\cr
L_0 v_P&=\bar L_0 v_P=(Q^2/4+P^2)v_P\cr
}\eqno(5.5)
$$
(so that the energy of $v_P$ is $2P^2-1/12$) and its descendants generated
by the action of $L_n$ and $\bar L_n$ with $n<0$ on $v_P$. Right and left
generators $L_n$ and $\bar L_n$ commute and therefore ${\cal A}_P$ has the
structure of a direct product of right and left modules.

To get more of an idea about ${\cal A}$ take the ``zero-mode'' of
the Liouville field $\phi(x)$
$$
\phi_0=\int_0^{2\pi}\phi(x){dx_1\over 2\pi}
\eqno(5.6)
$$
and consider the region $\phi_0\to-\infty$ in the configuration space. Here
one can neglect the exponential interaction term in the LFT action and
consider $\phi(x)$ as a free massless field which can be expanded as usual
in the free field oscillators
$$
\phi(x)=\phi_0-{\cal P}(z-\bar z)+\sum_{n\ne 0}\left({ia_n\over n}e^{inz}+
{i\bar a_n\over n}e^{-in\bar z}\right)
\eqno(5.7)
$$
Here
$$
{\cal P}=-{i\over 2}{\partial\over\partial\phi_0}
\eqno(5.8)
$$
is the momentum conjugate to the zero-mode and
$$
\eqalign{
\left[a_m,a_n\right]&={m\over 2}\delta_{m+n}\cr
\left[\bar a_m,\bar a_n\right]&={m\over 2}\delta_{m+n}\cr
}\eqno(5.9)
$$
The Virasoro generators are represented at $\phi_0\to-\infty$ as follows
$$
\eqalign{
L_n&=\sum_{k\ne 0,n}a_k a_{n-k}+(2{\cal P}+inQ)a_n\ \ \ \ \ \ \ n\ne 0\cr
L_0&=2\sum_{k>0}a_{-k}a_k+Q^2/4+{\cal P}^2\cr
}\eqno(5.10)
$$
and the same for $\bar L_n$ with $a_n$ substituted by $\bar a_n$. These
operators act in the space of states
$$
{\cal A}_0={\cal L}_2(-\infty<\phi_0<\infty)\otimes{\cal F}
\eqno(5.11)
$$
where ${\cal F}$ is the Fock space of the oscillators $a_n$, $\bar a_n$ with
$n\in Z$, $n\ne 0$. The space ${\cal F}$ consists of the Fock vacuum
$\left|0\right>$ annihilated by all $a_n$, $\bar a_n$ with $n>0$ and the
states generated by the action $a_n$, $\bar a_n$ with $n<0$ on $\left|0
\right>$.

Let $\Psi_s[\phi(x_1)]$ be the wave functional of any state $s\in{\cal A}$. The
$\phi_0\to-\infty$ asymptotic of this wave functional is naturally associated
with an element of (5.11). From (5.10) one observes that the zero-mode plane
wave $\exp(2iP\phi_0)$ times the Fock vacuum $\left|0\right>$ behaves under
(5.11) as the primary state $v_P$. It is clear that the correct wave functional
$\Psi_{v_P}[\phi(x_1)]$ of $v_P$ contains at $\phi_0\to-\infty$ also a
reflected wave $\exp(-2iP\phi_0)\left|0\right>$, i.e.,
$$
\Psi_{v_P}[\phi(x_1)]\sim\left(e^{2iP\phi_0}+S(P)e^{-2iP\phi_0}\right)\left|0
\right>\ \ \ \ {\rm at}\ \ \ \phi_0\to-\infty
\eqno(5.12)
$$
with some reflection amplitude $S(P)$ dependent on a more complicated
dynamics in the region of $\phi_0$ where the exponential interaction term
is important (see [11]). More generally, if $s_P$ is some state of ${\cal A}_P$
$$
\Psi_{s_P}[\phi(x_1)]\sim\left(e^{2iP\phi_0}+\hat S(P)e^{-2iP\phi_0}\right)
\left|s\right>,\ \ \ \ \phi_0\to-\infty
\eqno(5.13)
$$
where $\left|s\right>\in {\cal F}$ and $\hat S(P)$ is now a unitary operator
in ${\cal F}$. In particular $S(P)$ is the eigenvalue of $\hat S(P)$ on the
Fock vacuum $\left|0\right>$.

It is important to note that at a given $P$ all the matrix elements of the
operator $\hat S(P)$ are in fact determined by the conformal symmetry of LFT
up to an overall multiplier. For example, applying $L_{-1}$ in the form (5.10)
to the wave functional (5.12) we obtain
$$
\hat S(P)a_{-1}\left|0\right>=S(P){Q-2iP\over Q+2iP}a_{-1}\left|0\right>
\eqno(5.14)
$$
For more complicated Fock states the calculations are more involved but
always reduce to linear algebra and permit us to restore $\hat S(P)$
uniquely up to $S(P)$.

The following features are readily established. First, ${\cal F}$ is a tensor
product of the right and left modules (generated by $a_n$ and $\bar a_n$
respectively) and
$$
\hat S(P)=S(P)\hat s_R(P)\otimes\hat s_L(P)
\eqno(5.15)
$$
where $\hat s_R(P)$ and $\hat s_L(P)$ act independently in the right and
left Fock spaces, $\hat s_L(P)$ being isomorphic to $\hat s_L(P)$ under
$a_n\to\bar a_n$. Next, $\hat s_R(P)$ commutes with $L_0$ and therefore act
invariantly at each level (i.e., at every eigenspace of $L_0$). Moreover,
it can be argued that $\hat s_R(P)$ commutes with the infinite series of
the ``quantum KdV integrals of motion'' [22,23] built of the higher powers
of the stress tensor. In other words, it has the same eigenvectors as the
quantum KdV transfer matrix constructed in ref.[24].

However, generally $\hat s_R(P)$ is not known in a closed form. We quote here
its matrix elements at the second level spanned by the Fock states $\left|1,1
\right>=a_{-1}^2\left|0\right>$ and $\left|2\right>=a_{-2}\left|0\right>$
$$
\eqalign{
D(P)\hat
s_R(P)\left|1,1\right>&=\left(8P^3+(6Q^2-2)P+iQ(2Q^2+1)\right)\left|1,1
\right>+4iPQ\left|2\right>\cr
D(P)\hat s_R(P)\left|2\right>&=-8iPQ\left|1,1\right>+\left(-8P^3-(6Q^2-2)P+
iQ(2Q^2+1)\right)\left|2\right>\cr
}\eqno(5.16)
$$
where
$$
D(P)=\left(2P-iQ\right)\left(2P-i(b+2/b)\right)\left(2P-i(2b+1/b)\right)
\eqno(5.17)
$$

The overall factor $S(P)$ in (5.15)
is not prescribed by the conformal symmetry and has
to be recovered separately. It is easy to evaluate it in the semiclassical
limit $b\to 0$. Suppose that $P$ is also small of the order of $O(b)$. Then
even in the region of $\phi_0$ where the interaction is significant one can
neglect the oscillators in (5.7) and study only the dynamics of the zero-mode
$\phi_0$. In this approximation, known as the minisuperspace approach [20]
in LFT, the Hamiltonian (5.3) is substituted by
$$
H_0=-{1\over 12}-{1\over 2}{\partial^2\over\partial\phi_0^2}+2\pi\mu
e^{2b\phi_0}
\eqno(5.18)
$$
and the reflection amplitude (5.12) appears as
$$
S(P)=-\left({\pi\mu\over b^2}\right)^{-2iP/b}{\Gamma\left(1+2iP/b\right)\over
\Gamma\left(1-2iP/b\right)},\ \ \ \ b\to 0
\eqno(5.19)
$$
Comparing this with (3.18) we find it natural to propose the expression (3.18)
as the exact reflection amplitude $S(P)$ in LFT. In the next section we verify
this suggestion numerically using the thermodynamic Bethe ansatz technique.

\vskip 0.5cm

{\bf 6. Thermodynamic Bethe ansatz}

In this section we consider the sinh-Gordon model on a circle of circumference
$R$ with periodic boundary conditions. The problem is defined by the action
$$
A_{\rm shG}=\int dx_2\int_0^Rdx_1\left[{1\over 4\pi}\left(\partial_a\phi
\right)^2+2\mu\cosh b\phi\right]
\eqno(6.1)
$$
where $\phi(x_1,x_2)=\phi(x_1+R,x_2)$ is the periodic scalar field, $\mu\sim
[{\rm mass}]^{2+2b^2}$ is the dimensional coupling constant and $b$ is the
dimensionless parameter of the model. Due to the scaling properties of the
interaction operator $\exp(2b\phi)+\exp(-2b\phi)$ one can rescale the problem
to the circle of circumference $2\pi$ substituting (6.1) by
$$
S_{\rm shG}=\int dx_2\int_0^{2\pi}dx_1\left[{1\over 4\pi}\left(\partial_a\phi
\right)^2+\mu\left({R\over 2\pi}\right)^{2+2b^2}\left(e^{2b\phi}+e^{-2b\phi}
\right)\right]
\eqno(6.2)
$$

We are interested in the ground state energy $E(R)$ or, more conveniently, the
finite-size effective central charge
$$
c_{\rm eff}(R)=-{6R\over\pi}E(R)
\eqno(6.3)
$$
in the ultraviolet limit $R\to 0$. Let $\Psi_0[\phi(x_1)]$ be the ground state
wave functional and define again the zero-mode $\phi_0$ as in eq.(5.6). At
$R\to 0$ there is a large region $\log\mu(R/2\pi)^{2+2b^2}<2b\phi_0<-\log\mu
(R/2\pi)^{2+2b^2}$ in the configuration space where one can neglect the
interaction term in (6.2) and consider $\phi(x)$ as a free massless field
(5.7). In this region the ground state wave functional is expected to be a
superposition
$$
\Psi_0[\phi(x_1)]\sim\left(c_1 e^{2iP\phi_0}+c_2 e^{-2iP\phi_0}\right)
\left|0\right>
\eqno(6.4)
$$
of two zero-mode plane waves with some $R$-dependent zero-mode momentum $P$
times the Fock vacuum $\left|0\right>$ of the oscillators in (5.6).
The corresponding effective central charge is determined at $R\to 0$ mainly
by $P(R)$
$$
c_{\rm eff}(R)=1-24P^2+O(R^2)
\eqno(6.5)
$$
up to power corrections in $R$. The momentum $P(R)$ is quantized due to the
right and left potential walls at $2b\phi_0\sim\pm\log\mu(R/2\pi)^{2+2b^2}$ in
the action (6.2). Consider say the right wall at $2b\phi_0\sim-\log\mu
(R/2\pi)^{2+2b^2}$. If $R$ is small enough the second exponent $\exp(-2b\phi)$
in the potential term of (6.2) is small in this region and does not affect the
dynamics which is therefore expected to be essentially the same as in LFT.
Thus for the reflected wave we have
$$
\Psi_0[\phi(x_1)]\sim\left(e^{2iP\phi_0}+(R/2\pi)^{-4iPQ}S(P)e^{-2iP\phi_0}
\right)\left|0\right>
\eqno(6.6)
$$
where $Q$ is again given by eq.(2.3) and $S(P)$ is the same reflection
amplitude (3.18) as in LFT (5.12). Factor $(R/2\pi)^{-4iPQ}$ appears due to
the rescaling of $\mu$ in (6.2). A similar consideration about the left
wall reflection leads to the following quantization condition
$$
\left(R/2\pi\right)^{-8iPQ}S^2(P)=1
\eqno(6.7)
$$
For the ground state momentum this equation reads
$$
\delta(P)=\pi+4PQ\log(R/2\pi)
\eqno(6.8)
$$
where we have introduced the reflection phase $\delta(P)$
$$
S(P)=-\exp(i\delta(P))
\eqno(6.9)
$$
Together with (6.5) equation (6.8) determines the most important part of the
$R\to 0$ asymptotic of $c_{\rm eff}(R)$. For example, using the regular
expansion of the reflection phase in the odd powers of $P$
$$
\delta(P)=\delta_1(b)P+\delta_3(b)P^3+\delta_5(b)P^5+\ldots
\eqno(6.10)
$$
one can develop $c_{\rm eff}(R)$ systematically in $1/\log R$
$$
c_{\rm eff}(R)=1-{24\pi^2\over l^2}+{48\pi^4\delta_3(b)\over l^5}+\ldots
\eqno(6.11)
$$
where we have denoted
$$
l=\delta_1(b)-4Q\log(R/2\pi)
\eqno(6.12)
$$

On the other hand the sinh-Gordon model is integrable and its factorizable
scattering matrix is known [25]. This allows us to compute the same
effective central charge $c_{\rm eff}(R)$ by the thermodynamic Bethe ansatz
(TBA) technique [26,27]. In the TBA framework it is evaluated as the integral
$$
c_{\rm eff}(R)={3mR\over\pi^2}\int\cosh\theta\log\left(1+e^{-\varepsilon(
\theta)}\right)d\theta
\eqno(6.13)
$$
where $\varepsilon(\theta)$ is the solution to the following non-linear
integral equation
$$
mR\cosh\theta=\varepsilon(\theta)+\int\varphi(\theta-\theta')\log\left(1+
e^{-\varepsilon(\theta')}\right){d\theta'\over 2\pi}
\eqno(6.14)
$$
In (6.13) and (6.14) $m$ is the mass of the physical particle in the
sinh-Gordon spectrum. It is related to the coupling constant $\mu$ as [28]
$$
-{\pi\mu\over\gamma(-b^2)}=\left[{m\over 4\sqrt{\pi}}\Gamma\left({1\over 2+2b^2
}\right)\Gamma\left(1+{b^2\over 2+2b^2}\right)\right]^{2+2b^2}
\eqno(6.15)
$$
The kernel $\varphi(\theta)$ in eq.(6.14) contains the information about the
sinh-Gordon scattering and reads explicitly
$$
\varphi(\theta)={\displaystyle 4\sin{\pi b^2\over 1+b^2}\cosh\theta\over
\displaystyle \cosh 2\theta-\cos{2\pi b^2\over 1+b^2}}
\eqno(6.16)
$$
It is straightforward to solve eq.(6.14) numerically. At $R$ small enough
eqs.(6.5) and (6.8) can be interpreted as a parametric representation of the
``experimental'' reflection phase $\delta^{\rm (TBA)}(P)$
$$
\eqalign{
P&=\sqrt{1-c_{\rm eff}(R)\over 24}\cr
\delta^{\rm (TBA)}&=\pi+4PQ\log(R/2\pi)\ \ ,\cr
}\eqno(6.17)
$$
$R$ being the parameter. According to (6.5) we expect $\delta^{\rm (TBA)}(P)$
to reproduce the Liouville phase $\delta(P)$ up to exponentially small in
$1/P$ corrections
$$
\delta(P)=\delta^{\rm (TBA)}(P)+O\left(\exp\left(-{\pi\over 2PQ}\right)\right)
\eqno(6.18)
$$
In particular in the expansion
$$
\delta^{\rm (TBA)}(P)=\delta^{\rm (TBA)}_1(b)P+\delta^{\rm (TBA)}_3(b)P^3+
\delta^{\rm (TBA)}_5(b)P^5+\ldots
\eqno(6.19)
$$
all the coefficients are the same as in eq.(6.10).

We have solved eq.(6.14) numerically at different values of the parameter $b$
and estimated the few first coefficients $\delta^{\rm (TBA)}_{2k+1}(b)$
of (6.19). In Table 1 the numbers are compared with the corresponding
$\delta_{2k+1}(b)$ in the expansion (6.10) of the exact Liouville
amplitude (3.18)
$$
\eqalign{
\delta_1(b)&={4\over b}\log b^2-4Q\log{\displaystyle m\Gamma\left({1\over
2+2b^2}\right)\Gamma\left(1+{b^2\over 2+2b^2}\right)\over 4\sqrt{\pi}}+C\cr
\delta_3(b)&={16\over 3}\zeta(3)(b^3+b^{-3})\cr
\delta_5(b)&={64\over 5}\zeta(5)(b^5+b^{-5})\cr
}\eqno(6.20)
$$
etc. Here $C$ is the Euler's constant and in $\delta_1(b)$ the cosmological
constant $\mu$ is substituted in terms of $m$ by means of eq.(6.15). In the
numerical calculations we set $m=1$.

We consider the content of Table 1 as impressive evidence in support of the
exact Liouville reflection amplitude suggested in sect.3. The same TBA
analysis can be applied also for LFT with $c_L<25$. Parameter $b$ is complex
in this case and the sinh-Gordon scattering theory has to be replaced by the
staircase model [29]. We hope to say more about this interesting relation
in a future publication.

The exact Liouville reflection amplitude can be used in the opposite direction
in the analysis of the sinh-Gordon (or the staircase) model itself.
Subtracting the leading $R\to 0$ asymptotic of $c_{\rm eff}(R)$ predicted by
eqs.(6.5) and (6.8) from the TBA numerical data one can separate the
power-like corrections in (6.5). Perhaps this would allow us to clarify their
nature. Work along this line is now in progress.

\vskip 0.5cm

{\bf 7. Conformal bootstrap}

In this section we study numerically the conformal bootstrap equations
(2.21) using the representation (2.23) of the four-point function in LFT.
The structure constants $C(\alpha_1,\alpha_2,\alpha_3)$ are proposed
explicitly in sect.3. The conformal block ${\cal F}(\Delta_{\alpha_i},
\Delta,x)$ which also enters eq.(2.23) is not known generally in an
analytic form. In refs.[19] the following convenient representation has
been derived
$$
\eqalign{
{\cal F}(\Delta_{\alpha_i},\Delta,x)&=(16q)^{P^2}x^{Q^2/4-\Delta_{\alpha_1}-
\Delta_{\alpha_2}}(1-x)^{Q^2/4-\Delta_{\alpha_1}-\Delta_{\alpha_3}}\times\cr
&\phantom{Xx}
\left[\theta_3(q)\right]^{3Q^2-4(\Delta_{\alpha_1}+\Delta_{\alpha_2}+
\Delta_{\alpha_3}+\Delta_{\alpha_4})}H(\lambda_i^2,\Delta|q)\cr
}\eqno(7.1)
$$
Here
$$
\theta_3(q)=\sum_{n=-\infty}^\infty q^{n^2}
\eqno(7.2)
$$
and
$$
q=e^{i\pi\tau}
\eqno(7.3)
$$
is related to $x$ by the equation
$$
\tau=i{K(1-x)\over K(x)}
\eqno(7.4)
$$
where
$$
K(x)={1\over 2}\int_0^1 {dt\over\sqrt{t(1-t)(1-xt)}}
\eqno(7.5)
$$
Function $H(\lambda_i^2,\Delta|q)$ is better parameterized in terms of the
variables
$$
\lambda_i^2={Q^2\over 4}-\Delta_{\alpha_i}=\left({Q\over 2}-\alpha_i\right)^2
\eqno(7.6)
$$
instead of the external dimensions $\Delta_{\alpha_i}$. It satisfies the
following recurrence relation
$$
H(\lambda_i^2,\Delta|q)=1+\sum_{m,n}{q^{mn}R_{m,n}(\lambda_i)\over\Delta-
\Delta_{m,n}}H(\lambda_i^2,\Delta_{m,n}+mn|q)
\eqno(7.7)
$$
Here the sum is over all pairs $(m,n)$ of positive integers and
$$
\Delta_{m,n}={Q^2\over 4}-{(m/b+nb)^2\over 4}
\eqno(7.8)
$$
are the dimensions of degenerate representations of the Virasoro algebra
with the central charge (2.5). With the notation
$$
\lambda_{m,n}=(m/b+nb)/2
\eqno(7.9)
$$
the multipliers $R_{m,n}(\lambda_i)$ read explicitly
$$
R_{m,n}(\lambda_i)=2\ {\prod_{r,s}(\lambda_1+\lambda_2-\lambda_{r,s})
(\lambda_1-
\lambda_2-\lambda_{r,s})(\lambda_3+\lambda_4-\lambda_{r,s})(\lambda_3-
\lambda_4-\lambda_{r,s})\over\prod'_{k,l}\lambda_{k,l}}
\eqno(7.10)
$$
The products in (7.10) are over the following sets of integers $(r,s)$ and
$(k,l)$
$$
\eqalign{
r&=-m+1,-m+3,\ldots,m-3,m-1\cr
s&=-n+1,-n+3,\ldots,n-3,n-1\cr
}\eqno(7.11)
$$
and
$$
\eqalign{
k&=-m+1,-m+2,\ldots,m-1,m\cr
l&=-n+1,-n+2,\ldots,n-1,n\cr
}\eqno(7.12)
$$
while the prime sign near the last product symbol $\prod'_{k,l}$ means that
the two pairs $(k,l)=(0,0)$ and $(m,n)$ are missing. From the numerical point
of view the expression (7.7) is ``almost analytic''
in the sense that it permits
us to compute $H(\lambda_i^2,\Delta|q)$ very fast and precisely.

In view of eq.(7.1) we find it more convenient to write down the four-point
function (2.20) as
$$
\eqalign{
G_{\alpha_1,\alpha_2,\alpha_3,\alpha_4}(x,\bar x)&=(x\bar x)^{Q^2/4-
\Delta_{\alpha_1}-\Delta_{\alpha_2}}\left[(1-x)(1-\bar x)\right]^{Q^2/4-
\Delta_{\alpha_1}-\Delta_{\alpha_3}}\times\cr
&\phantom{Xx}
\left[\theta_3(q)\theta_3(\bar q)\right]^{3Q^2-4\sum_i\Delta_{\alpha_i}}
g_{\alpha_1,\alpha_2,\alpha_3,\alpha_4}(x,\bar x)\cr
}\eqno(7.13)
$$
and study the crossing properties of the reduced function
$$
\eqalign{
&g_{\alpha_1,\alpha_2,\alpha_3,\alpha_4}(\tau,\bar\tau)=\cr
&{1\over 2}\int C(\alpha_1,\alpha_2,Q/2+iP)C(\alpha_3,\alpha_4,Q/2-iP)
\left|(16q)^{P^2}H(\lambda_i^2,Q^2/4+P^2|q)\right|^2 dP\cr
}\eqno(7.14)
$$
{}From eqs.(2.21) it follows that
$$
\eqalignno{
g_{\alpha_1,\alpha_2,\alpha_3,\alpha_4}(\tau,\bar\tau)&=g_{\alpha_1,\alpha_2,
\alpha_4,\alpha_3}(\tau+1,\bar\tau+1)&(7.15a)\cr
g_{\alpha_1,\alpha_2,\alpha_3,\alpha_4}(\tau,\bar\tau)&=\left|\tau\right|^{3Q^2-
4\sum_i\Delta_{\alpha_i}}g_{\alpha_1,\alpha_3,\alpha_2,\alpha_4}(-1/\tau,-1/
\bar\tau)&(7.15b)\cr
}
$$
The first equation (7.15a) is identically satisfied by (7.14) due to the
following property of the function $H(\lambda_i^2,\Delta|q)$
$$
H(\lambda_1^2,\lambda_2^2,\lambda_3^2,\lambda_4^2,\Delta|q)=
H(\lambda_1^2,\lambda_2^2,\lambda_4^2,\lambda_3^2,\Delta|-q)
\eqno(7.16)
$$
which is easily derived from the relation (7.7) and eq.(7.10). Equation (7.15b)
still remains a non-trivial condition for the structure constants which is
believed to contain the whole information about the associativity of the
operator algebra.

For numerical calculations we have chosen the correlation function of four
puncture operators (2.11)\footnote{$^3$}{The corresponding values of
$\alpha_1, \alpha_2, \alpha_3, \alpha_4$ are well within the domain
(2.25) and so no ``discrete terms'' are expected to
appear in (2.23) and (7.14).}
$$
g(\tau,\bar\tau)={1\over 16}\left.
{\partial^4 g_{\alpha_1,\alpha_2,\alpha_3,\alpha_4}
(\tau,\bar\tau)\over\partial\alpha_1\partial\alpha_2\partial\alpha_3\partial
\alpha_4}\right|_{\alpha_1=\alpha_2=\alpha_3=\alpha_4=Q/2}
\eqno(7.17)
$$
After separating some irrelevant overall factors
$$
g(\tau,\bar\tau)=
{\Upsilon_0^8\over\pi^2}\left[\pi\mu\gamma(b^2)b^{2-2 b^2}\right]^{
-Q/b}\left({\tau-\bar\tau\over 2i}\right)^{-Q^2/2}f(\tau,\bar\tau)
\eqno(7.18)
$$
we have to verify the relation
$$
f(\tau,\bar\tau)=f(-1/\tau,-1/{\bar\tau})
\eqno(7.19)
$$
for the function
$$
f(\tau,\bar\tau)={1\over 2}\left({\tau-\bar\tau\over 2i}\right)^{Q^2/2}
\int r(P)\left|(16q)^{P^2}H(0,Q^2/4+P^2|q)\right|^2 dP
\eqno(7.20)
$$
Here
$$
\eqalign{
r(P)&={\pi^2 \Upsilon(2iP)\Upsilon(-2iP)\over\Upsilon_0^2\Upsilon^8(Q/2+iP)}\cr
&=\sinh{2\pi P\over b}\sinh(2\pi b P)\exp\left[-8\int_0^\infty{dt\over t}
{\sin^2Pt\left(1-e^{-Qt}\cos^2Pt\right)\over\sinh(bt)\sinh(t/b)}\right]\cr
}\eqno(7.21)
$$
We have computed $f(\tau,\bar\tau)$ for several values of $b$ and found that
eq.(7.19) is satisfied up to high numerical accuracy. In figs.1 and 2 functions
$f(\tau,\bar\tau)$ and $f(-1/\tau,-1/\bar\tau)$ are compared for
purely imaginary $\tau=it$ at
$b=0.8$ and $b=(1+i)/\sqrt{2}$. This last complex value corresponds to
$c_L=13$.

\vskip 2.0cm

{\bf 8. Accessory parameters}

The classical Liouville action (2.34) with $n$ singular points is closely
related to the classic problem of uniformization of Riemann surfaces [30]
and in particular to
the so-called problem of accessory parameters [31]. The last is basically
formulated as follows. Consider the ordinary linear differential equation
$$
\partial^2\psi(z)+\sum_{i=1}^n\left({1\over 4(z-x_i)^2}+{C_i\over z-x_i}
\right)\psi(z)=0
\eqno(8.1)
$$
with $n$ regular parabolic singular points $x_i$. The complex infinity
$z=\infty$ is supposed to be a regular point of (8.1) so that the
accessory parameters $C_i$ are restricted by the relations
$$
\eqalign{
\sum_{i=1}^n C_i&=0\cr
\sum_{i=1}^n (x_iC_i+1/4)&=0\cr
\sum_{i=1}^n (x_i^2 C_i+x_i/2)&=0\cr
}\eqno(8.2)
$$
The problem is to tune these parameters in such a way that the monodromy group
of eq.(8.1) is a Fuchsian one. It was proven in ref.[32] that the problem is
solved by the $\eta_i\to 1/2$ version of the Liouville action (2.34). In the
case $\eta_i\to 1/2$ one should be more careful since the boundary conditions
(2.33) for the Liouville equation (2.30) are slightly more complicated
$$
\eqalign{
\varphi(z,\bar z)&=-2\log|z|^2+O(1)\ \ \ \ \ \ \ \ \ \ \ \ \ \ \ \
\hbox{at}\ \ \ \ |z|\to\infty\cr
\varphi(z,\bar z)&=-\log|z-x_i|^2-2\log\log|z-x_i|^2+O(1)\ \ \hbox{at}\ \ \ \
|z-x_i|\to 0\cr
}\eqno(8.3)
$$
and more subtractions are needed to regularize the Liouville action (see
ref.[32] for more details). If $S^{\rm (cl)}(x_1,\ldots,x_n)$ is the
classical Liouville action of the solution $\varphi(z,\bar z|x_1,\ldots,x_n)$
to (2.29) with the boundary conditions (8.3) then the accessory parameters
$$
C_i=-{\partial\over\partial x_i}S^{\rm (cl)}(x_1,\ldots,x_n)
\eqno(8.4)
$$
solve the above problem.

{}From the LFT point of view the action $S^{\rm (cl)}(x_1,\ldots,x_n)$ can be
considered as the leading classical asymptotic $b\to 0$ of the $n$-point
function
$$
{1\over 2^n}\left.
{\partial^n{\cal G}_{\alpha_1,\ldots,\alpha_n}(x_1,\ldots,x_n)\over
\partial\alpha_1\ldots\partial\alpha_n}\right|_{\alpha_1=\ldots=\alpha_n=Q/2}
\sim\exp\left(-{1\over b^2}S^{\rm (cl)}(x_1,\ldots,x_n)\right)
\eqno(8.5)
$$
of the puncture operators (2.11). In this section we use the representation
(2.23) of the four-point function to get some information about $S^{\rm (cl)}
(x_1,\ldots,x_n)$ and therefore about the accessory parameters in the case
$n=4$. Note that due to (8.2) in this case there is only one independent
accessory parameter $C(x,\bar x)$ and eq.(8.1) can be reduced to
$$
\partial^2\psi(z)+\left({1\over 4z^2(1-z)^2}+{1\over 4(z-x)^2}+{x(1-x)
C(x,\bar x)\over z(1-z)(z-x)}\right)\psi(z)=0
\eqno(8.6)
$$
where $x$ is defined in eq.(2.19). Eq.(8.5) is reduced to
$$
{1\over 16}\left.
{\partial^4 G_{\alpha_1,\alpha_2,\alpha_3,\alpha_4}(x,\bar x)\over
\partial\alpha_1\partial\alpha_2\partial\alpha_3\partial\alpha_4}
\right|_{\alpha_1=\alpha_2=\alpha_3=\alpha_4=Q/2}\sim\exp\left(-{1\over b^2}
S^{\rm (cl)}(x,\bar x)\right)
\eqno(8.7)
$$
while
$$
C(x,\bar x)=-{\partial S^{\rm (cl)}(x,\bar x)\over\partial x}
\eqno(8.8)
$$

Let us now consider the classical limit $b\to 0$ of eq.(2.23) with four
``heavy'' operators $\alpha_i=\eta_i/b$. It is convenient to rescale the
integration variable $P=p/b$. The structure constants behave as in eq.(3.20)
in this limit while the conformal block ${\cal F}(\Delta_{\alpha_i},\Delta,x)$
has a similar $b\to 0$ asymptotic
$$
{\cal F}(\Delta_{\alpha_i},\Delta,x)\sim\exp\left({1\over b^2}f(\eta_i,p,x)
\right)
\eqno(8.9)
$$
Function $f(\eta_i,p,x)$ (which is sometimes called the classical conformal
block) is again generically unknown in a closed form but can be
straightforwardly developed in a power series in $x$
$$
f(\eta_i,p,x)=(\delta-\delta_1-\delta_2)\log x+{(\delta+\delta_1-\delta_2)
(\delta+\delta_3-\delta_4)\over 2\delta}x+O(x^2)
\eqno(8.10)
$$
where we have used the notations $\delta=p^2+1/4$ and $\delta_i=\eta_i(1-
\eta_i)$. At $b\to 0$ the integral in eq.(2.23) is determined by a saddle
point $p=p_s$, i.e., by the minimum of the function
$$
\eqalign{
{\cal S}_{\eta_1,\eta_2,\eta_3,\eta_4}(p|x,\bar x)&=S^{\rm (cl)}(\eta_1,
\eta_2,1/2+ip)
+S^{\rm (cl)}(\eta_3,\eta_4,1/2-ip)-\cr
&\phantom{XXXXXXX}f(\eta_i,p,x)-f(\eta_i,p,\bar x)\cr
}\eqno(8.11)
$$
where $S^{\rm (cl)}(\eta_1,\eta_2,\eta_3)$ is given explicitly by eq.(3.21).
Therefore the classical asymptotic of the four-point function is
$$
G_{\alpha_1/b,\ldots,\alpha_4/b}(x,\bar x)\sim\exp\left(-{1\over b^2}S^{\rm
(cl)}_{\eta_1,\ldots,\eta_4}(x,\bar x)\right)
\eqno(8.12)
$$
where
$$
S^{\rm (cl)}_{\eta_1,\eta_2,\eta_3,\eta_4}(x,\bar x)=
{\cal S}_{\eta_1,\eta_2,\eta_3,\eta_4}(p_s|x,\bar x)
\eqno(8.13)
$$
and $p_s$ is determined by the equation
$$
{\partial\over\partial p}{\cal S}_{\eta_1,\eta_2,\eta_3,\eta_4}(p|x,\bar x)=0
\eqno(8.14)
$$
With the exact formula (3.21) it reads explicitly
$$
2\pi-i\log S_{\eta_1,\eta_2}(p)-i\log S_{\eta_3,\eta_4}(p)=-{\partial
\over\partial p}\left(f(\eta_i,p,x)+f(\eta_i,p,\bar x)\right)
\eqno(8.15)
$$
where
$$
S_{\eta_1,\eta_2}(p)=
{\Gamma^2(1-2ip)\gamma(\eta_1+\eta_2-1/2+ip)\gamma(1/2+\eta_1-\eta_2+ip)\over
\Gamma^2(1+2ip)\gamma(\eta_1+\eta_2-1/2-ip)\gamma(1/2+\eta_1-\eta_2-ip)}
\eqno(8.16)
$$

The accessory parameter (8.8) corresponds to the special case $\eta_1=\eta_2=
\eta_3=\eta_4=1/2$ of eqs.(8.13--16)
$$
C(x,\bar x)={\partial\over\partial x}f(1/2,p_s,x)
\eqno(8.17)
$$
At small $x$ one can keep only the leading $x\to 0$ term in the classical block
(8.10) so that
$$
xC(x,\bar x)=(p_s^2-1/4)\left(1+O(x)\right)
\eqno(8.18)
$$
where $p_s$ is determined (up to power corrections in $x$) by the equation
$$
p\log x\bar x+4i\log{\Gamma(1-2ip)\Gamma^2(1/2+ip)\over\Gamma(1+2ip)
\Gamma^2(1/2-ip)}=\pi
\eqno(8.19)
$$
One also can systematically pick up the power corrections in $x$ using the
expansion
$$
\eqalign{
f(1/2,p,x)=&\left(p^2-{1\over 4}\right)\log x+\left(p^2+{1\over
4}\right){x\over
2}+\cr
&\left({13p^2\over 16}+{9\over 32}+{1\over 256(p^2+1)}\right){x^2\over 4}+
\left({23p^2\over 24}+{19\over 48}+{1\over 128(p^2+1)}\right){x^3\over 8}
+\ldots\cr
}\eqno(8.20)
$$

\vskip 0.5cm

{\bf 9. Conclusion}

The expression (3.14) for the three-point function, together with the
supporting evidence in sects.4--8, is the main result of this paper.
Although (3.14) is a conjecture we find the evidence convincing enough
to take it as the starting point in addressing some intriguing questions
in 2D Quantum Gravity.

The structure constants (3.14) allow one to construct (in principle)
the multipoint correlation functions of LFT through the decompositions
similar to (2.23). This gives access
to multipoint correlation functions of Quantum Gravity (coupled to a
matter theory) at fixed conformal moduli; this is in contrast with
the integrated correlation functions considered in [11--15]. This moduli
dependence of the correlation functions is expected to add some insight
on the nature of physical states in 2D Quantum Gravity.

One can also try to analyze the physics of non-minimal CFT (perhaps
non-unitary one to maintain $c_M < 1$, to begin with) or
non-conformal matter QFT coupled to Quantum Gravity.

The most interesting question is exactly what is it that happens to 2D
Quantum Gravity when the central charge $c_M$ of matter theory ($c_M =
26 - c_L$) exceeds 1. We hope that the structure constants (3.14) could
be an appropriate vehicle to enter this still rather mysterious
domain.

And of course it is important to extend the above analysis to
incorporate SUSY.

We hope to return to this questions in future.

\vskip 0.5cm

{\bf Acknowledgments\hfill}

One of the authors (Al.Z) is grateful to V.Fateev, P.Fendley and H.Saleur
for valuable discussions. A.Z. is grateful to A.Polyakov and S.Shenker
for interest to this work and discussion. Research of A.Z. is supported
by DOE grant \#DE-FG05-90ER40559.

\vskip 0.5cm

{\bf Note Added\hfill}

After this work was completed we have learned that the conjecture (3.14)
already exists in the literature. In ref.[34,35] the expression for the
Liouville three-point function apparantly equivalent to (3.14) is
proposed with the motivations very similar to those discussed in sect.3.
We are grateful to H.Dorn for bringing the papers [34,35] to our attention.

\vfill
\eject

\hbox{}
\vskip 1cm
\centerline{\bf References}
\vskip 8pt

\parskip 4pt
\parindent 35pt
\hsize=5.8in

\hskip -48pt 1. {\narrower
A.Polyakov. Phys.Lett., B103 (1981) 207.
\smallskip}

\hskip -48pt 2. {\narrower
T.Curtright and C.Thorn. Phys.Rev.Lett., 48 (1982) 1309; E.Braaten,
T.Curt\-right and C.Thorn. Phys.Lett., B118 (1982) 115; Ann.Phys., 147
(1983) 365.
\smallskip}

\hskip -48pt 3. {\narrower
J.-L.Gervais and A.Neveu. Nucl.Phys., B238 (1984) 125; B238 (1984) 396;
B257[FS14] (1985) 59.
\smallskip}

\hskip -48pt 4. {\narrower
E.D'Hoker and R.Jackiw. Phys.Rev., D26 (1982) 3517.
\smallskip}

\hskip -48pt 5. {\narrower
V.Kazakov. Phys.Lett., 150 (1985) 282; F.David. Nucl.Phys., B257 (1985) 45;
V.Kazakov, I.Kostov and A.Migdal. Phys.Lett., 157 (1985) 295.
\smallskip}

\hskip -48pt 6. {\narrower
E.Br\'ezin and V.Kazakov. Phys.Lett., B236 (1990) 144; M.Douglas and
S.Shen\-ker. Nucl.Phys., B335 (1990) 635; D.Gross and A.Migdal. Phys.Rev.Lett.,
64 (1990) 127.
\smallskip}

\hskip -48pt 7. {\narrower
V.Knizhnik, A.Polyakov and A.Zamolodchikov. Mod.Phys.Lett., A3 (1988) 819.
\smallskip}

\hskip -48pt 8. {\narrower
F.David. Mod.Phys.Lett., A3 (1988) 1651.
\smallskip}

\hskip -48pt 9. {\narrower
J.Distler and H.Kawai. Nucl.Phys., B321 (1989) 509.
\smallskip}

\hskip -53pt 10. {\narrower
M.Bershadsky and I.Klebanov. Phys.Rev.Lett.65(1990)3088.
\smallskip}

\hskip -53pt 11. {\narrower
J.Polchinski. Remarks on Liouville Field theory, in
``Strings 90'', R.Arnowitt et al, eds, World Scientific, 1991;
Nucl.Phys. B357 (1991) 241.
\smallskip}

\hskip -53pt 12. {\narrower
M.Goulian and M.Li. Phys.Rev.Lett., 66 (1991) 2051.
\smallskip}

\hskip -53pt 13. {\narrower
A.Polyakov. Mod.Phys.Lett. A6 (1991) 635.
\smallskip}

\hskip -53pt 14. {\narrower
P.Di Francesco and D.Kutasov. Phys.Lett. B261 (1991) 385.
\smallskip}

\hskip -53pt 15. {\narrower
Vl.Dotsenko. Mod.Phys.Lett. A6 (1991) 3601.
\smallskip}

\hskip -53pt 16. {\narrower
Vl.Dotsenko and V.Fateev. Phys.Lett. 154B (1985) 291.
\smallskip}

\hskip -53pt 17. {\narrower
V.Fateev and A.Zamolodchikov. Sov.Phys.JETP 82(1985)215;
Sov.J.Nucl.Phys. 43 (1986)657.
\smallskip}

\hskip -53pt 18. {\narrower
A.Belavin, A.Polyakov and A.Zamolodchikov. Nucl.Phys. B241 (1984) 333.
\smallskip}

\hskip -53pt 19. {\narrower
Al.Zamolodchikov. Commun.Math.Phys., 96 (1984) 419; Theor.Math.Phys., 73
(1987) 1088.
\smallskip}

\hskip -53pt 20. {\narrower
N.Seiberg. Notes on Quantum Liouville Theory and Quantum Gravity, in
``Random Surfaces and Quantum Gravity'', ed. O.Alvarez, E.Marinari,
P.Windey, Plenum Press, 1990.
\smallskip}

\hskip -53pt 21. {\narrower
Vl.Dotsenko and V.Fateev. Nucl.Phys., B251 (1985) 691.
\smallskip}

\hskip -53pt 22. {\narrower
R.Sasaki and I.Yamanaka. Adv.Stud. in Pure Math., 16 (1988) 271.
\smallskip}

\hskip -53pt 23. {\narrower
T.Eguchi and S.-K.Yang. Phys.Lett., B224 (1989) 373.
\smallskip}

\hskip -53pt 24. {\narrower
V.Bazhanov, S.Lukyanov and A.Zamolodchikov. Integrable structure of conformal
field theory, quantum KdV theory and thermodynamic Bethe ansatz. RU-94-98.
\smallskip}

\hskip -53pt 25. {\narrower
I.Arefyeva and V.Korepin. Pisma v ZhETF, 20 (1974) 680.
\smallskip}

\hskip -53pt 26. {\narrower
C.N.Yang and C.P.Yang. J.Math.Phys., 10 (1969) 1115.
\smallskip}

\hskip -53pt 27. {\narrower
Al.Zamolodchikov. Nucl.Phys., B342 (1990) 695.
\smallskip}

\hskip -53pt 28. {\narrower
Al.Zamolodchikov. Mass scale in sin-Gordon and its reductions. LPM-93-06.
\smallskip}

\hskip -53pt 29. {\narrower
Al.Zamolodchikov. Resonance factorized scattering and roaming trajectories.
ENS-LPS-335, 1991.
\smallskip}

\hskip -53pt 30. {\narrower
A.Poincare. J.Math.Pures Appl. (5) 4 (1898) 157.
\smallskip}

\hskip -53pt 31. {\narrower
F.Klein. Math.Ann. 21 (1883) 201; A.Poincare. Acta Math. 4 (1884) 201.
\smallskip}

\hskip -53pt 32. {\narrower
P.Zograf and L.Takhtajan. Functional Anal.Appl. 19 (1986) 219.
\smallskip}

\hskip -53pt 33. {\narrower
L.Takhtajan. Semi-Classical Liouville Theory, Complex Geometry of Moduli
Spaces, and Uniformization of Riemann Surfaces, in "New Symmetry
Principles in Quantum Field Theory", eds. J.Frolich et al, Plenum Press,
1992.
\smallskip}

\hskip -53pt 34. {\narrower
H.Dorn, H.-J.Otto. Phys. Lett. B291(1992) 39.
\smallskip}

\hskip -53pt 35. {\narrower
H.Dorn, H.-J.Otto. Nucl. Phys. B429 (1994) 375.
\smallskip}

\vfill
\eject

\hbox{}
\parindent 20pt
\parskip 12pt plus 1pt
\vskip 3cm
\centerline{\bf Figure Captions}
\vskip 4pt
\hskip -50pt Fig.1.
{\narrower
Reduced four-point function $f(\tau,\bar\tau)$ of eq.(7.20)
(continuous line) at $b=0.8$ and real $t=-i\tau$. It is compared with
$f(-1/\tau,-1/\bar\tau)$ (points) to verify the crossing symmetry
relation (7.19).
\smallskip}

\hskip -50pt Fig.2.
{\narrower
The same as in fig.1 but at $b=(1+i)/\sqrt{2}$ corresponding to $c_L=13$.
\smallskip}

\vfill
\eject

$\phantom{}${\bf Table 1.} First three coefficients $\delta^{\rm (TBA)}$ in
the expansion (6.19) obtained by numerical analysis the ``experimental''
reflection phase (6.17) in comparison with the corresponding ``exact'' ones
$\delta^{\rm (LFT)}$ given by eq.(6.20). The uncertainty in the
``experimental'' numbers is $\pm 1$ in the last (bracketed) digit.

\vskip 1.0cm

$$
\vbox{\tabskip=0pt \offinterlineskip
\halign to 410pt{\strut#& \vrule#\tabskip=1em plus2em& \hfil#\hfil&
                          \vrule#& \hfil#\hfil&
                          \vrule#& \hfil#\hfil&
                          \vrule#& \hfil#\hfil&
                          \vrule#& \hfil#\hfil&
                          \vrule#& \hfil#\hfil&
                          \vrule#& \hfil#\hfil&
\vrule# \tabskip=0pt\cr
\noalign{\hrule}
& &  & &  & &  & &  & &  & &  & &    & \cr
& & \omit\hidewidth ${b^2\over 1+b^2}$\hidewidth& &
    \omit\hidewidth $\delta_1^{\rm (TBA)}$\hidewidth& &
    \omit\hidewidth $\delta_1^{\rm (LFT)}$\hidewidth& &
    \omit\hidewidth $\delta_3^{\rm (TBA)}$\hidewidth& &
    \omit\hidewidth $\delta_3^{\rm (LFT)}$\hidewidth& &
    \omit\hidewidth $\delta_5^{\rm (TBA)}$\hidewidth& &
    \omit\hidewidth $\delta_5^{\rm (LFT)}$\hidewidth&  \cr
& &  & &  & &  & &  & &  & &  & &    & \cr
\noalign{\hrule}
& &  & &  & &  & &  & &  & &  & &    & \cr
& &$0.1$& &  & &$-16.61723$& &  & &$173.3336$& &  & &$-3225.315$& \cr
& &  & &  & &  & &  & &  & &  & &    & \cr
& &$0.2$& &$-4.74(4)$& &$-4.743972$& &$52.0(5)$& &$52.08913$&
&$-42(4).$& &$-425.1404$& \cr
& &  & &  & &  & &  & &  & &  & &    & \cr
& &$0.3$& &$-0.671(3)$& &$-0.6715864$& &$24.64(3)$& &$24.64884$&
&$-111.(2)$& &$-111.9785$& \cr
& &  & &  & &  & &  & &  & &  & &    & \cr
& &$0.4$& &$1.043(1)$& &$1.042998$& &$15.26(8)$& &$15.26739$&
&$-41.4(0)$& &$-41.39168$& \cr
& &  & &  & &  & &  & &  & &  & &    & \cr
& &$0.5$& &$1.533(5)$& &$1.533544$& &$12.81(9)$& &$12.82149$&
&$-26.5(6)$& &$-26.54535$& \cr
& &  & &  & &  & &  & &  & &  & &    & \cr
\noalign{\hrule}}}
$$

\vfill
\eject
\end